\documentstyle[epsf,psfig]{aipproc}

\newcommand{\Ha}{H$\alpha$}

\newcommand{\Hb}{H$\beta$}
\newcommand{\Ms}{$M_{\odot}$}

\newcommand{\lya}{Ly$\alpha$}

\def\ltsima{$\; \buildrel < \over \sim \;$}
\def\simlt{\lower.5ex\hbox{\ltsima}}
\def\gtsima{$\; \buildrel > \over \sim \;$}
\def\simgt{\lower.5ex\hbox{\gtsima}}

\begin{document}

\title{The Spectra of Star Forming Galaxies at High Redshift}

\author{Max Pettini$^1$, Charles C. Steidel$^2$, Mark Dickinson$^3$,
Melinda Kellogg$^2$, Mauro Giavalisco$^4$, Kurt L. Adelberger$^2$}

\address{
$^1$ Royal Greenwich Observatory, Madingley Road, Cambridge CB3 0EZ, 
England\\
$^2$ Palomar Observatory, Caltech 105$-$24, Pasadena, CA 91125, USA\\
$^3$ Department of Physics and Astronomy, The Johns Hopkins University,
Baltimore, \\
MD 21218, USA\\
$^4$ The Carnegie Observatories, 813 Santa Barbara Street, Pasadena, 
CA 91101, USA
}

\maketitle

\begin{abstract}
We review the spectral properties of the population of star forming 
galaxies at $z \sim 3$ discovered using a colour selection 
technique which targets the Lyman 
discontinuity at 912~\AA. 
The UV luminosity of the typical 
$z \sim 3$ galaxy exceeds
by more than one order of magnitude 
that of the most luminous starbursts
in the nearby universe, although the maximum star formation 
intensity ({\it SFR} per unit area) is within the limits found in local 
surveys.   
We consider in detail the likely magnitude of dust extinction and 
conclude that published estimates of the volume-averaged star formation 
rate at high $z$ need to be revised to higher values by a factor of about 
3. This correction improves the agreement between the observations and 
recent theoretical predictions.
\lya\ emission is generally weak, most likely as a result of resonant 
scattering. 
The large equivalent widths of the strongest 
interstellar lines 
and their systematic blueshift (by up to several 
hundred km~s$^{-1}$) relative to the \lya\ emission line
are indicative of highly energetic outflows
in the interstellar medium. 
Pilot observations have detected the redshifted \Hb\ and [O~III] emission 
lines in the $K$ band. The widths of these features imply dynamical masses of 
$\approx 10^{10}$~\Ms\ for the innermost star forming regions;
the total masses involved are likely to exceed $10^{12}$~\Ms.
\end{abstract}

\section*{Introduction}

The technique of Lyman limit imaging has proved to be highly 
successful in identifying star forming galaxies 
at high redshifts \cite{sph95}. 
At the time of writing (June 1997) our survey covers 
some 800 square arcminutes of sky in a dozen fields, with approximately 
600 candidates brighter than  
${\cal R} \simeq 25.5$ of which 230 
are spectroscopically confirmed galaxies 
at $2 < z < 4$\,.
The most significant results of this programme have been described by Mauro 
Giavalisco at this conference; here I review 
what we have learnt in the past year on the physical properties of 
galaxies at these early epochs from consideration of their spectra.
In the spirit of the meeting, I shall emphasize the similarities 
(and differences) between
the spectra of high redshift galaxies and those of
local starbursts; such a comparison draws extensively 
on the detailed studies of
local starbursts with the {\it Hubble Space Telescope} being carried out
by Tim Heckman, Claus Leitherer, Gerhardt Meurer and their collaborators
and described elsewhere in this volume.
 
I shall focus in particular on: {\it (a)} the UV luminosities and implied star 
formation rates; {\it (b)} evidence for the presence of dust and the 
corresponding UV extinction;  
{\it (c)} \lya\ emission and large-scale velocity fields;
and {\it (d)} the prospects for detecting the familiar emission lines 
from H~{\sc II} regions and the key role they play in clarifying and 
extending much of the information provided by the ultraviolet spectra.
Although the bulk of the spectra
are of only modest signal-to-noise ratio,
they are nevertheless adequate for addressing items 
{\it (a), (b)} and {\it (c)}.  
For point {\it (d)} on the other hand, and for detailed studies of the 
stellar and interstellar lines which are not discussed here for lack of 
space, we rely
on the most luminous objects in the sample to which 
we have paid special attention and 
devoted long exposure times.
The best case is  
the $z = 2.723$ galaxy 1512-cB58 \cite{yee96}.
This spectrum (reproduced in Figure 1) 
is of comparable, or better, quality than {\it HST} spectra of local 
starburst galaxies and is a veritable mine of information for
detailed studies of the physical conditions in high redshift galaxies.
It now seems highly likely that this object, which is
$\sim 4$ magnitudes brighter than the typical $z \simeq 3$ galaxy in our 
sample, is not extraordinarily luminous but gravitationally lensed
\cite{wl97,seitz97} and 
therefore presumably provides us with an unusually clear view of a `normal' 
galaxy at high redshift.

%
%
\begin{figure}[t]
\unitlength1cm
\begin{picture}(16.5,16.5)  
{\epsfxsize=16.5cm
\epsfbox{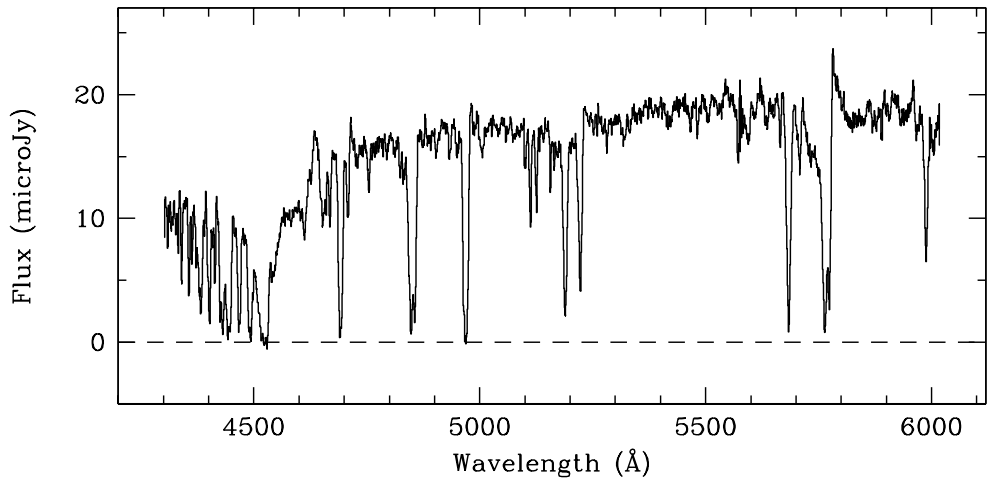}}
\end{picture}
\vskip -9cm
\caption{Spectrum of the $V = 20.64$, $z = 2.723$  
galaxy 1512-cB58 obtained with LRIS on the
Keck~I telescope in October 1996. With a total exposure time  
of 11\,400~s we reached S/N $\simeq 50$ per pixel at a resolution of 3.5 
\AA.}
\end{figure}

\section*{Ultraviolet Luminosities}

The typical high-$z$ galaxy in our survey, with 
${\cal R} = 24.5$, ($G-\cal{R}$) = 0.5, and $z = 3$,
has a far-UV luminosity
$L_{1500} = 1.3 \times 10^{41}~h_{70}^{-2}$
erg~s$^{-1}$~\AA$^{-1}$ at 
1500~\AA\ 
(unless otherwise stated, we adopt a
$H_0 = 70$~km~s$^{-1}$~Mpc$^{-1}$, $\Omega_M = 0.2$, $\Omega_{\Lambda} = 0$
cosmology throughout this article).
It is instructive to compare this value with those measured in 
nearby starbursts. 
The UV luminosities we find at $z = 3$ are
$\approx 800$ times greater than that of the brightest star cluster
in the irregular galaxy NGC~4214 studied with {\it HST} by 
Leitherer et al. 1996 \cite{leitherer96},
and exceed by a factor of $\approx 30$ that of the most luminous local 
example, the Wolf-Rayet galaxy NGC~1741 
which contains $\approx 10^4$ O type stars
\cite{conti96}.

The ultraviolet spectra, in which we see the integrated 
continuum of O and early B stars, 
can in principle be used to estimate the star formation rate in 
a more direct way than the Balmer lines, which are produced
by the reprocessed ionizing radiation of 
the stars at the very tip of the IMF.
Adopting a continuous star formation model with an age 
greater than $10^8$~years 
and a Salpeter IMF from 0.1 to 100~\Ms\ \cite{bc96,lrh95}, 
the typical $L_{1500} = 1.3 \times 10^{41}~h_{70}^{-2}$
erg~s$^{-1}$~\AA$^{-1}$ corresponds to 
a star formation rate 
{\it SFR} $\simeq 8~h_{70}^{-2}$~\Ms~yr$^{-1}$.
This is probably a lower limit, since dust extinction
(see below) and a lower age would both raise this value
(for an age of $10^7$~years the implied  
{\it SFR} is greater by a factor of $\approx 1.7$).

It is interesting to note that even the brightest objects in our sample
fall well within the surface brightness distribution of local starbursts.
The highest values of $L_{1500}$ which we have found 
are $\sim 4-5$ times higher 
than the median. 
Adopting a typical extinction correction of a factor of 
$ \approx 3$ at 1500~\AA\ (see below)
and a typical half-light radius
$r \simeq 2$~kpc \cite{mauro96},
we arrive at a star formation intensity
({\it SFR} per unit area) 
$\dot{\Sigma} \sim 13$~\Ms~yr$^{-1}$~kpc$^{-2}$\,.
This value compares well with the upper envelope of 
the ultraviolet sample 
considered by Meurer et al. 1997 \cite{meurer97}.
Thus, the star forming galaxies which we are finding at high 
redshift appear to be spatially more extended versions
of the local starburst phenomenon. 
The same physical processes which 
limit the maximum star formation intensity in nearby starbursts, as 
discussed by Meurer et al., also seem to be at play  
in young galaxies at high redshift
which may well be undergoing their first episodes of star formation.

\section*{Dust Extinction}

Dust is a ubiquitous component of the interstellar medium; 
given that galaxies at $z = 3$
are obviously already enriched in heavy elements, 
it is likely that some dust is mixed with the gas and stars
we observe.
Unfortunately, even relatively small amounts of dust can have a
significant effect in the rest-frame ultraviolet and thereby alter 
our view of high-$z$ galaxies. 
In particular, dust will: 
{\it (a)} extinguish  
resonantly scattered emission lines, most notably \lya;
{\it (b)} attenuate the UV continuum leading to underestimates of the
star formation rate; and
{\it (c)} redden the broad spectral energy distribution 
so that it resembles that of an older stellar population.
To some extent we have to learn to live with these
problems because of the inherent uncertainties of any dust corrections 
which arise mostly from the unknown shape of the extinction 
law.

The intrinsic slope of the integrated UV continuum 
of a star forming galaxy is a robust quantity which, as explained for 
example by Calzetti 1997a \cite{daniela97a}, varies little with 
the exact shape of the IMF or the age of the starburst.
Spectral synthesis models show that 
the continuum between 1800 and 1200 \AA\ is well 
approximated by a power law of the form $f_{\nu} \propto \lambda^{\alpha}$
(where $f_{\nu}$ is the flux per frequency unit) with
$\alpha$ between  $-0.5$ and 0. Similarly,
the empirical template starburst spectrum constructed by Calzetti 1997b
\cite{daniela97b} has $\alpha = -0.1$\,.
In contrast, the galaxies we observe generally have UV spectral slopes
between 0 and $+1.5$; 
the spectrum of cB58 reproduced in Figure 1, for example, can be clearly 
seen to be redder than flat spectrum and  
has $\alpha = +1.3$\,.
Such (relatively) red spectra can result from an aging starburst or from 
an IMF lacking in massive stars, but we regard both possibilities as 
unlikely because with sufficiently high S/N 
we see directly the spectral signatures of O stars.
The most straightforward interpretation, in analogy with local 
starbursts, is that the spectra are reddened by dust extinction.

In Figure 2, reproduced from
\cite{med97}, we use 
the ($G-\cal{R}$) colours of the entire sample
of spectroscopically confirmed galaxies 
to estimate dust corrections to the star 
formation rates at high redshift. 
Assuming an intrinsic spectral slope 
$\alpha = -0.13$, as is the case for a   
Bruzual \& Charlot 1996 \cite{bc96} model
with 1~Gyr old continuous star formation and Salpeter IMF,
the curves labelled with different values of 
$E(B-V)$ at the top of the figure show the predicted 
($G-\cal{R}$) colour as a function of redshift, if the spectra of high-$z$ 
galaxies are reddened with an extinction law 
similar to that which applies to  
stars in the Small Magellanic Cloud.
The curves all rise to redder ($G-\cal{R}$) colour
with redshift due to the increasing line blanketing by the 
\lya\ forest \cite{madau95}.

The difference between the observed and predicted ($G-\cal{R}$) colour
yields the extinction at 1500~\AA, $A_{1500}$,
appropriate to each galaxy; by adding together the individual 
values for all the galaxies in the sample
Dickinson et al. 1997 \cite{med97} 
deduce a net correction by a factor of 1.8
to the comoving volume-averaged star formation rate 
(in \Ms~yr$^{-1}$~Mpc$^{-3}$)
at $z = 3$\,. 
Adopting the greyer extinction law deduced by 
Calzetti et al. from
the integrated spectra of nearby starbursts \cite{cks94,daniela97a},
increases the net correction to a factor of 3.5\,.
Also shown in Figure 2 is the zero-reddening curve for 
a younger stellar population (10~Myr) which 
has a bluer intrinsic UV slope $\alpha = -0.42$ \cite{bc96}.
If the models are correct, some of the galaxies 
we have found are evidently younger than $10^9$ years, since they lie 
between the two zero-reddening curves in 
Figure 2.\footnote{There are a few points with ($G-\cal{R}$) colours {\it bluer}
than the $10^7$ years curve in Figure 2. 
\lya\ emission line contamination may play a role for some, particularly
for the most deviant object labelled ``G2'' which is an 
AGN with strong line emission.
Differences by $\simlt 0.1$~mag can easily be explained by photometric 
errors and the stochastic nature of the \lya\ forest 
blanketing.} Adopting the $10^7$ year old model as the unreddened 
template leads to corrections to the global star formation rate
by factors of 3.5 and 6.3, for the SMC and Calzetti et al. extinction 
laws respectively.  

%
%
\begin{figure}
\hspace*{-0.25in}
\psfig{figure=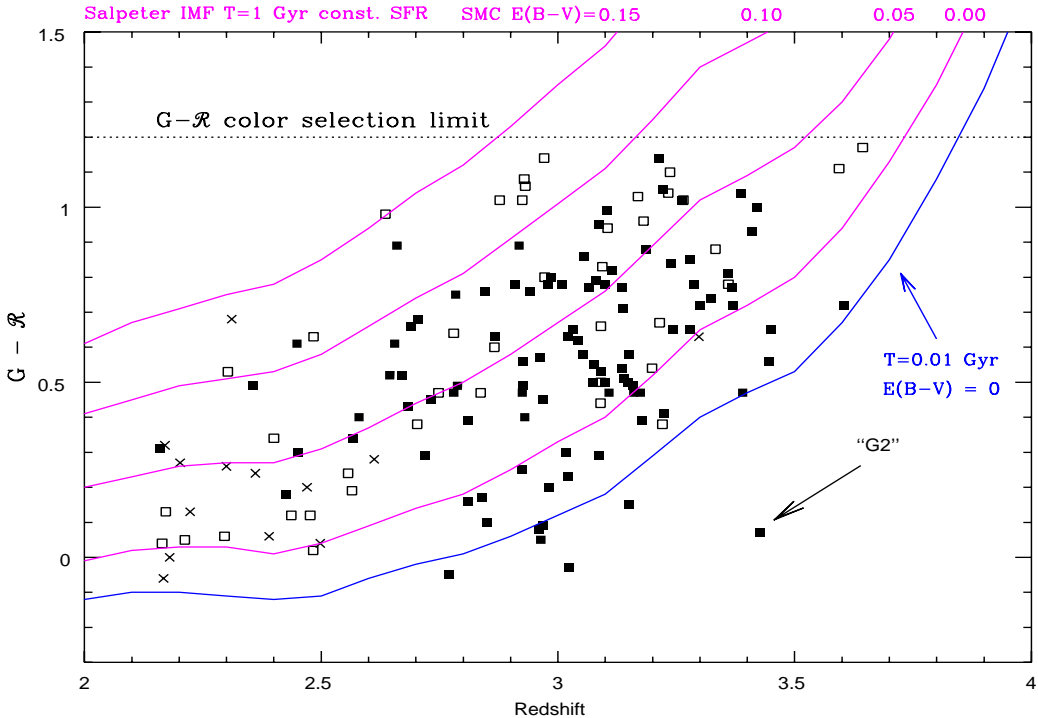,width=15.0cm,height=10.0cm,angle=270}
\hspace*{-0.75in}
\caption{Comparison between observed and predicted 
($G-\cal{R}$) colours of high-$z$ star forming galaxies 
for different amounts of SMC-type ultraviolet extinction and for 
different ages of the stellar populations. 
The symbols correspond to 
different photometric selection criteria, but all the galaxies plotted here 
have spectroscopically confirmed redshifts \protect\cite{med97}.}
\end{figure}

We conclude that the likely dust correction to the  
integrated ultraviolet luminosity of Lyman break galaxies 
at $z = 3$ amounts to a factor of $\approx 3$. 
The correction
could be as low as $\sim 2$ and as high as $\sim 6$, 
depending on the 
age of the stellar population and on the wavelength dependence of the 
ultraviolet extinction. 
Our estimate is $\sim 5$ times lower than that deduced by Meurer et 
al. 1997 \cite{meurer97}. The difference results from a combination of 
several factors, the most significant of which
is that the value $A_{1620} = 2.9$~mag (a factor of 15 in flux) proposed
by Meurer et al. refers to the most extreme case 
considered above, an age of $10^7$ years and the Calzetti et al. 
attenuation curve.
Given their number density, roughly comparable to that of present day 
$L^{\ast}$ galaxies, we consider it very unlikely that
all Lyman break galaxies are as young as $10^7$ years.
The lower values of UV extinction derived here 
are supported by the measured \Hb\ flux in the few cases where
this line has been detected in Lyman break galaxies (see below).

%
%
\begin{figure}
\vspace*{-1.5cm}
\hspace*{2.2cm}
\psfig{figure=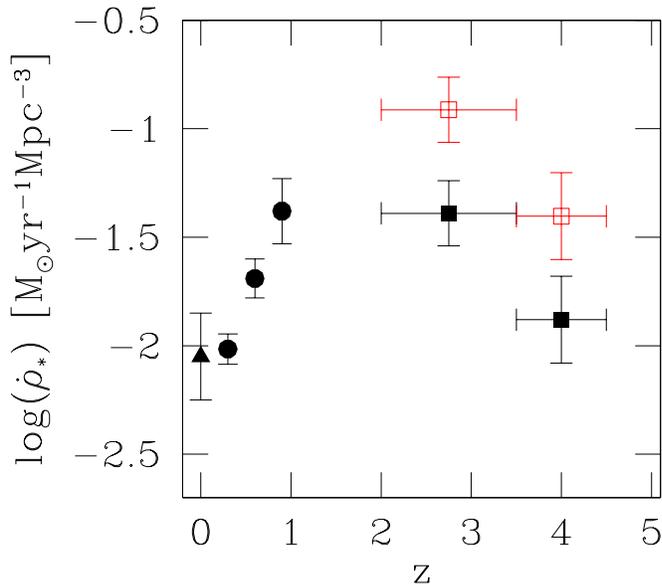,width=9.5cm,height=9.5cm,angle=0}
\hspace*{-0.75in}
\vspace*{0.5cm}
\caption{The comoving volume-averaged star formation rate as a function 
of redshift, reproduced from \protect\cite{madau97} for $H_0 = 50$~km~s$^{-1}$ and
$q_0 = 0.5$. The filled squares are measurements from the {\it HDF}, the circles 
from the {\it CFRS} and the triangle from a local \Ha\ survey. The open squares
show the {\it HDF} values with the correction 
we derive for dust extinction 
in Lyman break galaxies.}
\end{figure}

Figure 3 shows the effect of a factor of 3 correction for 
dust extinction on the comoving volume-averaged star formation rate 
deduced by Madau 1997 \cite{madau97} 
from the density of $U$ and $B$ drop-outs in the 
{\it Hubble Deep Field}. 
Given that the uncorrected value of $\dot{\rho}_{\ast}$ 
at any epoch deviates by only a factor of $\sim 3$ from the average
over the Hubble time, the inclusion of dust clearly has a significant impact on 
the interpretation of the cosmic star formation history.
The plot in Figure 3 is generally taken to indicate a peak 
in star formation between $z \sim 1$ and $\sim 2.5$.
This conclusion probably still holds once dust is taken into account, 
although    
the corrections appropriate to the lower redshift galaxies in the {\it 
CFRS} 
have yet to be determined \cite {cfrs95}. 
Since the values of $\dot{\rho}_{\ast}$ from the {\it CFRS} survey are based on 
galaxy luminosities in the near-UV (2800~\AA), 
the same amount of dust as deduced for the Lyman break galaxies would 
result in an upward correction of the filled circles in Figure 3 by a 
factor of $\sim 2$\,.
On the other hand, the typical $A_V \simeq 1$~mag found by
Tresse et al. 1996 \cite{tresse96} from the Balmer decrement of 
{\it CFRS} galaxies at $z \leq 0.3$ 
would imply a correction by a factor of $\sim 3$ to the UV continuum at 
2800 \AA\ \cite{daniela97a}.
It is interesting to note that the revisions we propose
to Madau's plot bring the observed values of $\dot{\rho}_{\ast}$
in good agreement with recent theoretical predictions based 
on cold dark matter models of galaxy formation
(see Figure 16 of Baugh et al. 1997 \cite{baugh97})
and on the chemical evolution of the Milky Way
\cite{ps97}.

\section*{\lya\ Emission}

The \lya\ emission line is detected in about 75\% of the galaxies in our 
sample but is always weaker than expected on the basis of the UV 
continuum luminosities, in agreement with the generally null results of 
previous searches for high redshift galaxies based on this spectral feature
(e.g. \cite {tdt95}).
There are strong indications that the main reason for the weakness of 
\lya\ emission is resonant scattering in an outflowing interstellar medium.
When detected, the emission line is generally redshifted by up to several 
hundred km~s$^{-1}$ relative to the interstellar absorption lines and, in 
the best observed cases, its profile is clearly 
asymmetric.

The bright galaxy 0000-D6, reproduced in Figure 4, is a good example.
The zero point of the velocity scale in Figure 4 is at 
$z = 2.960$, the redshift of 
the strong interstellar absorption lines in 0000-D6. 
The peak of \lya\ emission is 
at a relative velocity of 800 km~s$^{-1}$, and while the red wing 
extends to $\sim 1500$~km~s$^{-1}$, the blue wing is sharply 
absorbed. This P~Cygni-type profile can be understood as originating in an 
expanding envelope around the H~II region; 
the unabsorbed \lya\ photons we see are those
back-scattered from the receding part of the nebula. 
In agreement with this picture, we find that the
systemic velocity of 
the star-forming region is $\approx 400$~km~s$^{-1}$,
as measured from the wavelengths of 
weak photospheric lines from O stars 
(S~V~$\lambda 1501.96$ and O~IV~$\lambda 1343.35$),
which can be discerned in this high S/N spectrum, and of
[O~III]~$\lambda5007$ which we have detected in the $K$-band (see below).
Taken together, the relative velocities of interstellar, stellar and 
nebular lines point to large scale outflows in the interstellar medium, 
presumably as a consequence of the starburst activity
which in this galaxy, one of the brightest in our sample, 
approaches $\approx 100~h_{70}^{-2}$~\Ms~yr$^{-1}$.
Similar, although generally less energetic, outflows are seen in local 
starburst galaxies observed with {\it HST} and {\it HUT} 
\cite{kunth96,rosa97}.

%
%
\begin{figure}[t]
\unitlength1cm
\begin{picture}(16.5,16.5)  
{\epsfxsize=16.5cm
\epsfbox{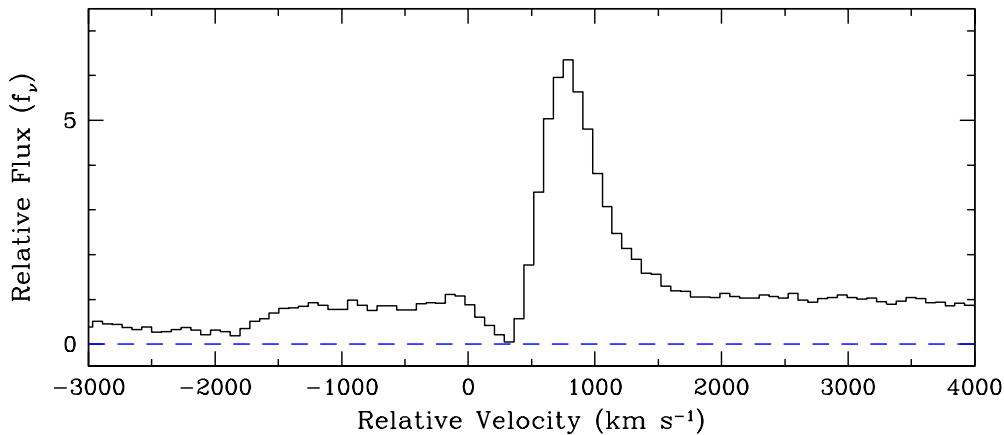}}
\end{picture}
\vskip -10cm
\caption{The wavelength region near \lya\ in the
${\cal R} = 22.9$, 
$z = 2.963$ galaxy 0000-D6.
The spectrum was obtained with LRIS on Keck I by Hy Spinrad and Arjun Dey
\protect\cite{spinrad97} with an  
exposure time of 17\,650~s and a resolution of
4~\AA\ FWHM. 
The equivalent width of the combined \lya\ emission and absorption
feature is 10.5 \AA.
}
\end{figure}

Large scale motions of the type we have found in 0000-D6 could be the 
main reason for the strengths of the interstellar absorption lines in 
high-$z$ galaxies which, with typical equivalent widths of 
2-3~\AA, are often greater than their counterparts in nearby starbursts.
(Since these lines are saturated, their equivalent widths are much more 
sensitive to the velocity dispersion of the gas than to the metallicity 
of the gas).
On the other hand, such strong absorption lines are also often
seen in damped \lya\ systems, which are not generally
associated with sites of active star formation and where they may
reflect the complex velocity fields of merging protogalactic clumps
\cite{hsr97}.

In any case, the systemic redshift of \lya\ emission relative to 
interstellar absorption {\it along the same sight-line} brings into 
question the validity of interpreting such differences along adjacent 
sight-lines as evidence for large rotating disks, as recently proposed by
Djorgovski et al. 1996, 
and Lu, Sargent, \& Barlow 1997 \cite{djorgovski96,lu97}.

\section*{Infrared Prospects}

The rest-frame optical spectrum of star forming galaxies is strikingly 
different from the ultraviolet (e.g. \cite{conti96}). 
At $z \sim 3$ the
nebular emission lines which dominate the optical spectrum are redshifted 
into the infrared $H$ and $K$ bands; there is a strong incentive to 
detect and measure these lines as they 
hold important clues to the nature of 
high-$z$ galaxies. 
In  particular: {\it (a)} the line widths, which presumably reflect the 
overall kinematics of the star forming regions in a galaxy,  
can provide an indication of the masses involved; 
{\it (b)} a detection of \Hb\ (or \Ha\ at $z \simlt 2.5$) 
would give a measure of the star formation rate which can be compared 
with that deduced from the UV continuum; and 
{\it (c)} the ratios of the familiar nebular lines are probably the 
most promising way of estimating the metallicity of these galaxies, 
given the complexity of the ultraviolet absorption line spectra.

Somewhat paradoxically (given that the discovery of $z \sim 3$
galaxies awaited the availability of large telescopes)
the detection of the strongest rest-frame 
optical emission lines in the $K$-band is 
actually within reach of 4-m telescopes equipped with
moderately high dispersion near-infrared spectrographs.
What is required is prior knowledge of the galaxy redshift and sufficient 
spectral resolution to ensure that the lines of interest fall in 
a gap between the much stronger OH$^-$ emission features which dominate 
the infrared sky.
Pilot observations which we carried out with the CGS4 spectrograph on UKIRT 
in September 1996 were successful in detecting \Hb\ and/or [O~III] 
emission lines in both $z \simeq 3$ galaxies targeted, 
0000-D6 and 0201-C6 \cite{pettini97a}; 
a third measurement---in cB58---has been obtained by 
Gillian Wright (private communication).
Figure 5 shows a portion of the $K$-band spectrum of  
0201-C6. \Hb\ and the weaker member [O~III] doublet are both detected at 
the $\sim 5 \sigma$ level, 
whereas the stronger [O~III] line is lost in the 
nearby sky emission.

%
%
\begin{figure}[t]
\unitlength1cm
\begin{picture}(16.5,16.5)  
{\epsfxsize=16.5cm
\epsfbox{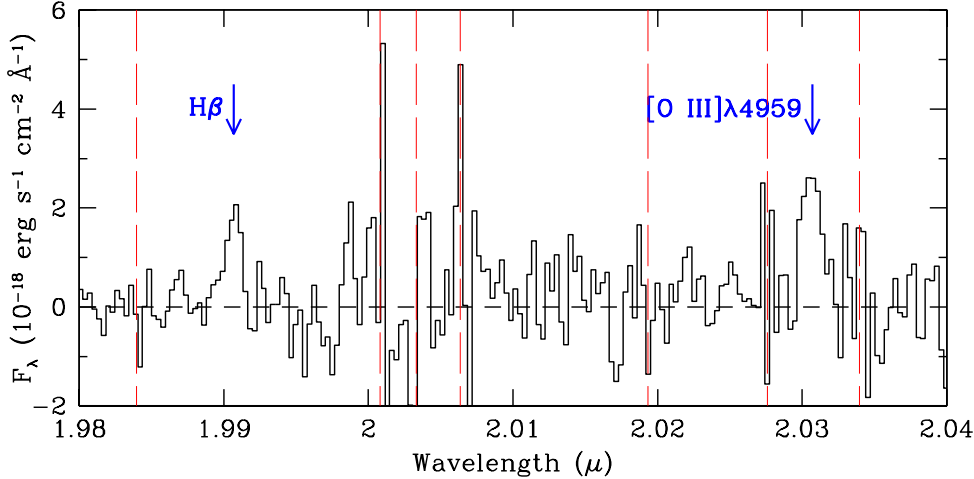}}
\end{picture}
\vskip -9cm
\caption{Portion of the infrared spectrum of the ${\cal R} = 23.9$, 
$z = 3.059$ galaxy 0201-C6 obtained with CGS4 on UKIRT
in September 1996 \protect\cite{pettini97a}. 
The exposure time was 18\,000~s and the resolution is
8~\AA\ FWHM sampled with 2.7 wavelength bins. 
The vertical dashed lines indicate the locations of the
major OH$^-$ sky emission lines.
}
\end{figure}

As can be seen from Figure 5, both \Hb\ and [O~III]
emission lines are resolved; after correcting for the instrumental 
resolution, we measure $\sigma = 70 \pm 20$~km~s$^{-1}$\,.
A similar velocity dispersion is also found in the other two cases, 
0000-D6 and cB58. 
(Incidentally, the fact that cB58 shows the same velocity dispersion 
as the other two $z \simeq 3$ galaxies, even though it is $\simgt 10$ 
times brighter, is another indication of its gravitationally lensed 
nature). 
If we combine $\sigma = 70$~km~s$^{-1}$ with the half-light radii
of $\approx 2$~kpc deduced for both 
0201-C6 and 0000-D6 from {\it HST} WFPC2 images, 
we obtain virial masses
of $\sim 1.2  \times 10^{10}$\Ms. 
This is comparable to the mass of the Milky Way bulge \cite{dwek95}
and to the dynamical mass within the central $r = 2$~kpc 
of an $L^{\ast}$ elliptical galaxy. However, 
the total masses involved are likely to be substantially greater,
given that the present 
IR observations sample only the innermost cores of the galaxies, 
where the star formation rates are presumably highest.
Indeed the clustering properties of 
Lyman break galaxies strongly suggest that they are associated 
with dark matter halos of mass $M \simgt 10^{12}$\Ms\ \cite{steidel97}.

The \Hb\ flux of 0201-C6 (Figure 5), 
$(2.6 \pm 0.6) \times 10^{-17}~h_{70}^{-2}$ erg~s$^{-1}$~cm$^{-2}$,
corresponds to a  luminosity
$L_{\rm {H}\beta} = (2.3 \pm 0.5 ) \times 10^{42}~h_{70}^{-2}$ erg~s$^{-1}$.
Adopting an \Ha/\Hb\ ratio of 2.75
and Kennicutt's \cite{kennicutt83}
calibration 
{\it SFR} (\Ms yr$^{-1}$) 
$= L_{\rm {H}\alpha}$ (erg~s$^{-1}$)/$1.12 \times 10^{41}$ 
which is appropriate for a Salpeter IMF from 0.1 to 100 \Ms, 
we deduce a star formation rate 
{\it SFR}$_{\rm {H}\beta} = (55 \pm 13)~h_{70}^{-2}$~\Ms\ yr$^{-1}$.
For comparison 
{\it SFR}$_{\rm{UV}} = (20 - 35)~h_{70}^{-2}$~\Ms yr$^{-1}$, 
depending on whether the age of the starburst is 
$10^9$ or $10^7$ years respectively.
Estimates of the star formation rate from \Ha\ emission and the UV continuum do 
not normally agree to better than a factor of $\sim 2$ in local starbursts
(e.g. \cite{meurer95}); the agreement in 0201-C6 is further improved when 
account is taken of the small amount of reddening ($E(B-V) \simlt 0.1$)
implied by the slope of the UV continuum ($\alpha = 0.35$) using the 
prescription by Calzetti \cite{daniela97a}.
We reach similar conclusions in 0000-D6 and cB58; in all three cases 
values of ultraviolet extinction as high as proposed by Meurer et al.
($A_{1620} \simeq 3$~mag \cite{meurer97}) predict higher \Hb\ fluxes than 
observed.

Finally, it is interesting to note that, with 
$L_{\rm {H}\beta} = (2.3 \pm 0.5 ) \times 10^{42}~h_{70}^{-2}$ erg~s$^{-1}$
and $\sigma = 70$~km~s$^{-1}$,
0201-C6 falls on the extrapolation to higher luminosities
of the correlation found for local H~II galaxies
by Melnick, Terlevich, \& Moles 1988 \cite{mtm88}.
 
These preliminary results demonstrate clearly the great potential of 
infrared observations for complementing the information provided by 
the rest-frame ultraviolet and ultimately 
leading to a better understanding of the 
nature of high-$z$ galaxies. 
With large telescopes
the detection of nebular emission lines in the near-IR will soon
become routine and it will be possible to address the points touched upon 
here in greater depth, using a large set of measurements.
Galaxies at redshifts such that [O~II]~$\lambda 3737$, 
[O~III]~$\lambda\lambda 4959, 5007$ and \Hb\ all fall within gaps between 
sky emission are the highest priority for future observations, because 
they offer the means to determine the abundance of oxygen.
Bringing together emission and absorption 
\cite{pettini97b} measurements at high redshift
will allow a more comprehensive 
description of the chemical enrichment history of the universe
than is possible at present.

\end{document}